\documentclass[journal, letterpaper]{IEEEtran}

\usepackage{graphicx}
\usepackage{url}        
\usepackage{amsmath}    
\usepackage{amssymb}
\usepackage{cite}
\usepackage{multirow}
\usepackage{array}
\usepackage{algorithm,algorithmic}
\usepackage{textcomp}
\usepackage[colorlinks=true,
            linkcolor=blue,    
            citecolor=magenta, 
            urlcolor=magenta,
            breaklinks=true   
]{hyperref}

\begin{document}

	\title{AI-Based Stroke Rehabilitation Domiciliary Assessment System with ST-GCN Attention}
	\author{Suhyeon Lim, Ye-eun Kim, and Andrew J. Choi.}
	\markboth{Arxiv Preprint}{}
	\maketitle

\begin{abstract}
Effective stroke recovery requires continuous rehabilitation integrated with daily living. To support this need, we propose a home-based rehabilitation exercise and feedback system. The system consists of (1) hardware setup with RGB-D camera and wearable sensors to capture stroke movements, (2) a mobile application for exercise guidance, and (3) an AI server for assessment and feedback. When a stroke user exercises following the application guidance, the system records skeleton sequences, which are then assessed by the deep learning model, RAST-G@ (Rehabilitation Assessment Spatio-Temporal Graph ATtention). The model employs a spatio-temporal graph convolutional network to extract skeletal features and integrates transformer-based temporal attention to figure out action quality. For system implementation, we constructed the NRC dataset, include 10 upper-limb activities of daily living (ADL) and 5 range-of-motion (ROM) collected from stroke and non-disabled participants, with Score annotations provided by licensed physiotherapists. Results on the KIMORE and NRC datasets show that RAST-G@ improves over baseline in terms of MAD, RMSE, and MAPE. Furthermore, the system provides user feedback that combines patient-centered assessment and monitoring. The results demonstrate that the proposed system offers a scalable approach for quantitative and consistent domiciliary rehabilitation assessment. Our code is available at
\href{https://github.com/LimSuH/NRC-rehab}{https://github.com/LimSuH/NRC-rehab}

\end{abstract}


\section{Introduction}
\label{sec:introduction}
\IEEEPARstart{R}{ECENT} advancements in neurology, particularly in motor control and learning, have revealed different mechanisms that induce changes in brain plasticity and behavior over both short- and long-term periods. Physical rehabilitation can be seen as a form of motor learning that occurs under specific conditions\cite{1,2,3,4}, and patients with motor impairments, such as those following a stroke, are capable of limited motor learning, although with variations in learning speed and volume. In particular, usage-based and reward-based learning, which are shaped by habitual, repetitive actions and rewards, play a key role in determining long-term brain and behavioral changes in stroke patients after they are discharged and resume daily activities.\cite{5} Therefore, rehabilitation performed at domestic environments is significant in the patient’s recovery. However, without expert, improper rehabilitation can lead to injury and a professional training is essential for evaluations and feedback in order to provide effective rehabilitation. In this study, we propose AI-based assessment model called RAST-G@ and a outline of rehabilitation monitoring system that can be used within the domiciliary environments. This system predicts the user’s rehabilitation movements RGB-D video data and expert assessment score collected during the system development and experiment. Many existing studies provide limited evaluation for topical motion segments such as range of angle. But ours takes the sequence of actions as input and learns it with a spatio-temporal graph based deep learning method. Since movements are continuous inputs through time, temporal feature learning is important to evaluate the quality of actions. We enhanced temporal learning results by introducing temporal attention, which has proven the performance in deep learning. Evaluation questionnaires from professional therapists contain the quality of complex actions. RAST-G@, the model that end-to-end deep learning-based rehabilitation assessment is able to assist healthcare providers and contributes to quantitative and consistent assessments of rehabilitation without the need for separate standardization processes.
In this paper, we propose an end-to-end deep learning-based rehabilitation assessment model. This model assists healthcare providers and contributes to quantitative and consistent assessments of rehabilitation without the need for separate standardization processes. It has the potential to enhance the activation of the rehabilitation and the quality of healthcare.

\section{Related Works}
Deep learning has applicated in rehabilitation research, as these methods enable the automated processing of complex motion data and biosignals, thereby reducing the subjectivity inherent in conventional clinical assessments and allowing for more precise and objective measurements. The integration of deep learning is expected to enhance therapeutic efficiency, support the development of personalized rehabilitation strategies, and facilitate continuous monitoring of functional recovery. Within this context, two primary types of tasks are generally distinguished: \textit{Recognition and Assessment.}

\subsection{Human Action Representation}
A high-quality dataset enable the extraction of more informative features. However, human behavior is abstract, which has motivated substantial research on effective data representations. Early work primarily relied on sensor-based modalities.\cite{6,7}
Sensor-based approaches are widely used in rehabilitation because they support continuous monitoring and are often cost-effective and robust to environmental variation when compared with RGB video. For instance, Panwar et al.\cite{8}, attached tri-axial accelerometers to classify upper-limb movements, including reaching and retrieving, lifting, and arm rotation in stroke rehabilitation. Pogorelc et al.\cite{9} similarly analyzed wearable sensor signals to identify gait patterns in older adults to facilitate early detection of health issues.

Despite these advantages, sensor-based methods depend on correct device placement and attachment. Modeling whole-body coordination may require additional processing or instrumentation. Vision-based modalities offer an alternative with simple setup. However, They often require more computation and can be raise privacy problems. They can also be sensitive to background noise.\cite{10}
To balance these trade-offs, recent deep learning studies increasingly adopt structured, processed representations, particularly skeleton-based data.\cite{7,11}. Skeleton data represent the body as joint landmarks connected according to anatomical topology. This representation naturally supports graph-based modeling of joint interactions.
It can also reduce sensitivity to background noise\cite{12}, relative to RGB-based approaches, while partially mitigating occlusion and depth ambiguity through the use of kinematic structure and temporal context.

\subsection{Recognition and Assessment}
Recognition tasks categorize sequential input data, to predefined discrete labels. In human action recognition, the goal is to identify and distinguish actions from motion trajectories or sensor-derived signals.
In contrast to classification tasks, recognition commonly operates on continuous streams that may include irrelevant segments or transitions, which necessitates temporal features and robustness to ambiguity.
A broad range of deep learning architectures has been developed for action recognition. Among them, the Spatio-Temporal Graph Convolutional Network called ST-GCN\cite{13} has shown strong performance on skeleton-based data. In a skeleton graph, joints are modeled as nodes and anatomical connections are represented as edges, which enables explicit modeling of coordinated joint motion. 
Because human actions are defined by structured interactions among joints, ST-GCN is suited to capturing relational dependencies. Building on spatio-temporal skeleton recognition, relative work has also reported that even simple augmentation strategies can yield state-of-the-art performance.\cite{14}
Duan et al.\cite{15} proposed an alternative representation that converts skeleton sequences into heatmaps and trains a 3D convolutional network. This approach reduces outliers from pose estimation errors and improves action recognition performance.

In rehabilitation, recognition aims to classify rehabilitation-specific movements. This is challenging because patient motion is often irregular, requiring identification and segmentation of meaningful behaviors and conversion into clinically interpretable outputs.
The assessment task is related to recognition because accurate evaluation requires clear discrimination of movements and extraction of meaningful features. In rehabilitation, assessment aims to quantify movement quality.
However, rehabilitation metrics are often subjective and can vary substantially across evaluation tools. To address this limitation, earlier studies adopted distance-based metrics to derive quantitative assessments.
Representative examples include approaches based on Mahalanobis distance\cite{16}  and Dynamic Time Warping called DTW\cite{17,18,19,20}, which compute dissimilarity between a reference movement and a patient’s movement. Although these methods yield objective values, they can be sensitive to noise and have limited validation as clinically reliable metrics. Capecci et al.\cite{21}. reported that DTW-based scores can diverge from clinician scores. 


Other work has mapped handcrafted features to clinical scores\cite{22}, but such pipelines often require multiple components and increase computational cost.
The Fugl–Meyer Assessment (FMA) is a widely adopted traditional rehabilitation instrument. Because it provides quantitative scores, it can also be considered a practical and digitization-friendly target metric for training AI-based assessment models. However, its three-level scoring scheme\textminus \{0: unable to perform, 1: partial performance, 2: full performance\}\textminus provides limited granularity. This scale requires detailed descriptions of motion and evaluation criteria, making it less suitable for automated assessment.

\begin{figure}[!h]
\centerline{\includegraphics[width=\columnwidth]{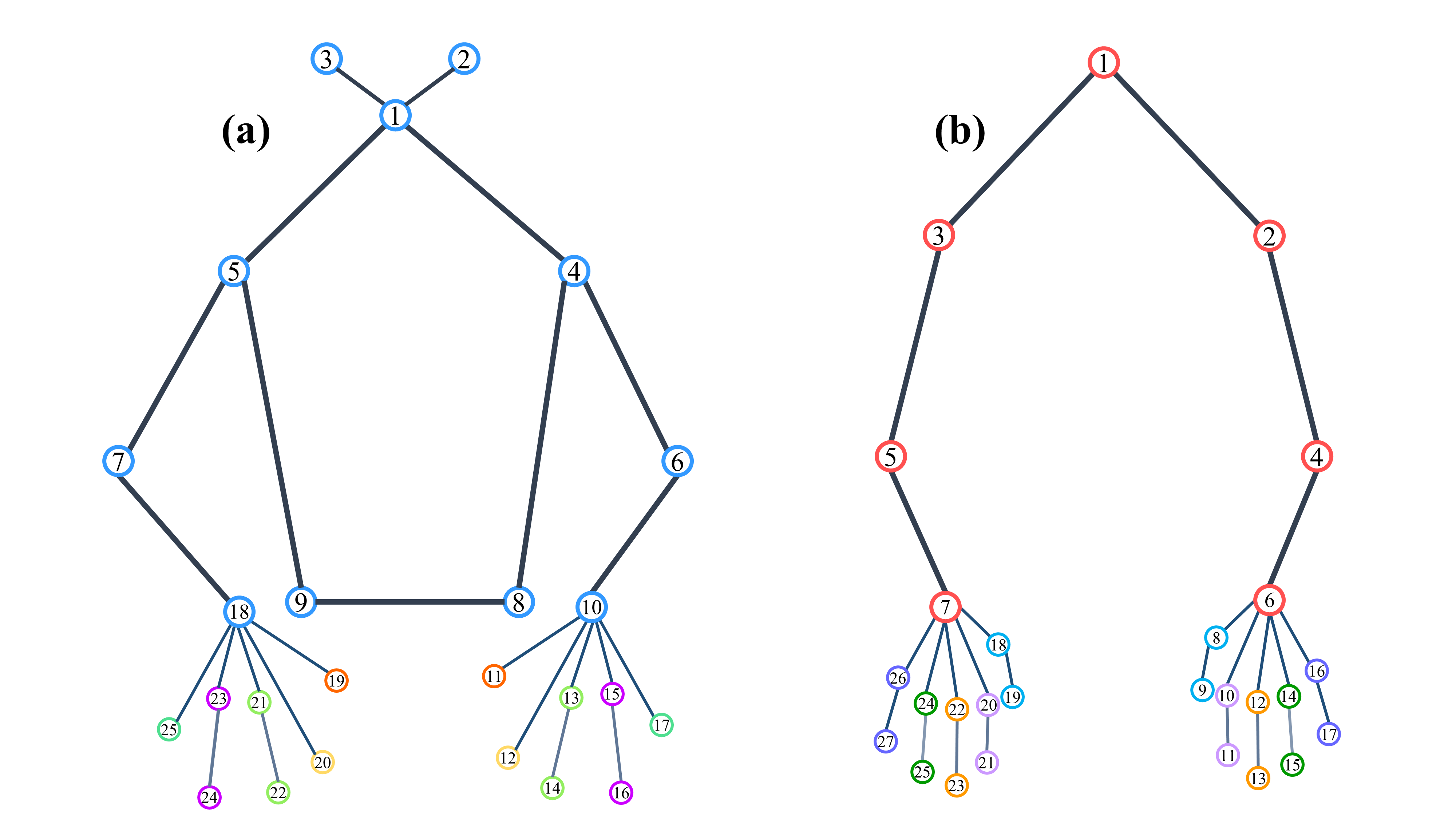}}
\caption{Upper limb skeleton keypoints used in this study. (a) Basic 25 configuration, (b) Additional 27 configurations.}
\label{fig1}
\end{figure}

\begin{figure}[t]
\centerline{\includegraphics[width=\linewidth]{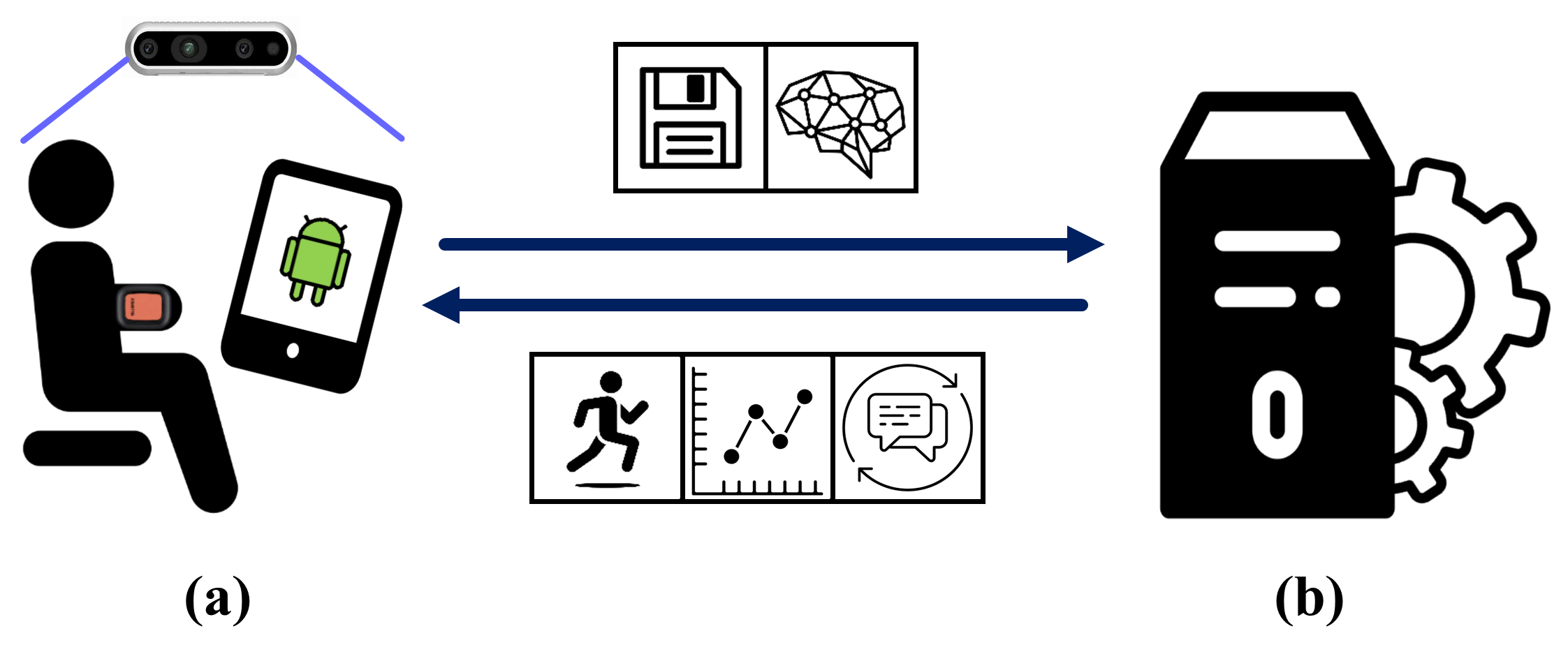}}
\caption{Overall system architecture. (a) Hardware setup and rehabilitation exercise interface using the proposed system (RGB-D camera, IMU, and android OS mobile device), (b) Server component for storing collected data, running model inference, and managing feedback transmission.}
\label{fig2}
\end{figure}

\subsection{Deep Learning Network for Human Action}
Traditionally, deep learning frameworks for human action have modeled motion by combining spatial encoders with temporal modeling modules. The temporal dimension is particularly important because the sequential ordering and dynamics of movement frequently influence performance more strongly than static pose configurations.
Liao et al.\cite{23} proposed a framework that jointly incorporates spatial structure and temporal dynamics. Their method progressively learned features from five body-part subsets and then integrated them to represent whole-body motion, thereby encouraging the model to capture temporal dependencies. 
Given the limited size of available rehabilitation datasets, they additionally employed a temporal pyramid strategy that subsamples sequences to improve learning efficiency and robustness. However, conventional convolutional networks used for spatial feature extraction are not inherently matched to skeleton data, where joints interdependence through anatomical connectivity\cite{24}.
This limitation has motivated the adoption of Graph Convolutional Networks called GCN, which preserve the anatomical connectivity of the human body during model learning\cite{25}.
Building on this direction, Deb et al.\cite{26} introduced GCN-based modeling into rehabilitation assessment. By learning sequential skeleton data with a ST-GCN, their approach enabled a single model to evaluate multiple action classes. 

Following earlier work, they also incorporated Long Short-Term Memory(LSTM) networks for temporal features, drawing inspiration from Liao et al.\cite{27} LSTMs are a variant of Recurrent Neural Networks(RNN) that use gating mechanisms to mitigate vanishing gradients and to retain information over longer horizons. Nevertheless, LSTM can still be limited in capturing very long-range dependencies, and their architectural complexity may increase the risk of overfitting and reduce interpretability.\cite{28}

\section{Methods}
The objective of this study is to learn a regression function $f_{\theta}: X \to Y$ that maps an input motion sequence $X \in \mathbb{R}^{N \times C \times T \times V \times M}$ to an assessment score $Y$, which is a scalar value provided by clinicians for both stroke patients and Non-Disabled (ND) participants. Here, $N$ is the batch size, $C$ is the channel dimension, $T$ is the sequence length, $M$ is the number of instances, and $V$ is the number of keypoints shown in Fig. 1(a).

The output $Y$ is a scalar score annotated by clinicians for both stroke and ND participants. Based on this formulation, we describe the acquisition of clinically relevant rehabilitation data and the design of assessment items, and the proposed deep learning network that combines ST-GCN blocks and transformer-based Temporal Attention modules to effectively capture the spatio-temporal characteristics of skeleton sequences.

\begin{table}[t]
\caption{NRC dataset overall}
\label{table0}
\centering
\setlength{\tabcolsep}{10pt}
{\renewcommand{\arraystretch}{1.25}
    \begin{tabular}{c |c c cc}
        \noalign{\hrule height 0.75pt}
        \hline
        {\centering Subject \par} & {\centering Train\par} & {\centering Validation\par} & {\centering Test\par} & {\centering Total\par}\\
        \hline
        ND & 293 & 32 & - & 325\\
        Stroke & 633 & 70 & 114 & 817\\ 
        \hline
        Total & 916 & 112 & 114 & 1,142\\
        \hline
        \noalign{\hrule height 0.75pt}
    \end{tabular}}
\end{table}

\subsection{Collecting Dataset}
Domiciliary rehabilitation systems require both computational efficiency and intuitive usability. To meet these requirements, we designed an end-to-end pipeline in which all modules operate in a coordinated manner. Because stroke rehabilitation data are costly and difficult to obtain, the software performs recognition and assessment whenever a user completes an exercise, while also recording the resulting data for model training. The overall system architecture is illustrated in Fig. 2.
The hardware configuration consists of a mobile tablet, an RGB-D camera, and IMU sensors. The software comprises a recording module and an assessment module. The recording module manages exercise execution and data acquisition, whereas the assessment module communicates with a deep learning assessment model on the server to compute and display the predicted score.

\subsubsection{\textbf{Rehabilitation Actions Data}}
Using the proposed system, we collected a multimodal rehabilitation dataset. The hardware setup included an Intel RealSense D435i depth camera and Movella Xsens Dot sensors. We captured RGB-D videos and collected accelerometer and gyroscope signals $(x,y,z)$ from sensors attached to both wrists. For the assessment model presented in this paper, we used only RGB-D data to maintain a compact and interpretable pipeline. As summarized in Table I, data were collected from two participant groups: ND participants and stroke patients.
Table II lists the 15 action classes included in the dataset, comprising Activities of Daily Living (ADL) and Range Of Motion (ROM) tasks selected to evaluate upper-limb functional ability.\cite{29} The movement set was finalized in consultation with faculty members in physical therapy.
For the model input, skeleton keypoints were estimated from the captured RGB-D videos. We adopted a 25-keypoint configuration, which is widely used in skeleton-based human analysis.\cite{ye-eun} This configuration supports compatibility with commonly used skeleton-based representations and benchmarks, including OpenPose\cite{30} and COCO WholeBody\cite{31}. It also facilitates direct performance comparisons with prior work and aligns with well-established graph construction strategies. This data collection study was approved by the IRB of Gachon University.(IRB no. 1044396-202306-HR-102-02)

\begin{table}[t]
\caption{15 ACTION CLASSES WITH 10 ADL, 5 ROM}
\label{table1}
\setlength{\tabcolsep}{2pt}
{\renewcommand{\arraystretch}{1.25}
\begin{tabular}{p{0.03\linewidth} p{0.28\linewidth} p{0.38\linewidth} p{0.07\linewidth} p{0.15\linewidth}}
\noalign{\hrule height 0.75pt}
\hline
\# & {\centering Class\par} & {\centering Description\par} & {\centering Cate.\par} & {\centering Tool\par} \\
\hline
 {\centering 1\par} & LiftCupHandle        & Drinking water with a cup    & {\centering UNI\par} & {\centering Cup\par} \\
 {\centering 2\par} & HairBrush            & Brushing hair                & {\centering UNI\par} & {\centering Comb\par} \\
 {\centering 3\par} & BrushTeeth           & Brushing teeth               & {\centering UNI\par} & {\centering Toothbrush\par} \\
 {\centering 4\par} & RemoteCon            & Remote controller            & {\centering UNI\par} & {\centering Controller\par} \\
 {\centering 5\par} & MovingCan            & Moving a tool                & {\centering UNI\par} & {\centering Can\par} \\
 {\vspace{0pt}\vfill \centering 6\vfill} & \vspace{0pt}\vfill \raggedright Writing \vfill & Writing with one hand\par while holding paper\par with the other hand&  {\vspace{0pt}\vfill \centering BIA\vfill} &  {\vspace{0pt}\vfill \centering Pencil\vfill} \\
 {\centering 7\par} & FoldingPaper         & Folding paper                & {\centering BIA\par} & {\centering Paper\par} \\
 {\centering 8\par} & FoldupTowel          & Folding towel                & {\centering BIA\par} & {\centering Towel\par} \\
 {\centering 9\par} & WashFace             & Washing face                 & {\centering BIS\par} &  {\centering -\par} \\
 {\centering 10\par} & Smartphone           & Using cellphone              & {\centering BIS\par} & {\centering Cellphone\par} \\
 {\centering 11\par} & RightShoulderFrontal & Flexion the right arm  & {\centering BIA\par} & {\centering -\par} \\
 {\centering 12\par} & LeftShoulderFrontal  & Flexion the left arm   & {\centering BIA\par} & {\centering -\par} \\
 {\centering 13\par} & RightShoulderSide    & Abduction the right arm & {\centering BIA\par} & {\centering -\par} \\
 {\centering 14\par} & LeftShoulderSide     & Abduction the left arm  & {\centering BIA\par} & {\centering -\par} \\
 {\centering 15\par} & LateralRotation      & Shoulder external rotation & {\centering BIS\par} & {\centering -\par} \\

\hline
\noalign{\hrule height 0.75pt}
\end{tabular}}
\label{table1}
\end{table}

\begin{table}[t]
\caption{Rehabilitation Exercise Assessment Questionnaires are Scored on 0-5 Likert Scale.}
\label{table1}
\setlength{\tabcolsep}{2.5pt}
{\renewcommand{\arraystretch}{1.25}
\begin{tabular}{p{0.05\linewidth} p{0.9\linewidth}}
\noalign{\hrule height 0.75pt}
\hline
\# & {\centering Questionnaire\par} \\
\hline
 {\centering 1\par} & Are the main objectives of the action achieved? \\
 {\centering 2\par} & Is the body moved without instability during the action? \\
 {\centering 3\par} & Is the action performed smoothly and continuously without interruption? \\
 {\centering 4\par} & Is the head movement performed correctly? \\
 {\centering 5\par} & Is the right arm performed correctly? \\
 {\centering 6\par} & Is the left arm performed correctly? \\
 {\centering 7\par} & Is the trunk performed correctly? \\
 {\centering 8\par} & Is the task performed without dropping or missing the tool in the hand? \\
 {\centering 9\par} & Are the force, speed, and direction appropriately controlled during the action? \\
 {\centering 10\par} & Is the action trajectory self-regulated during motion? \\
\hline
\noalign{\hrule height 0.75pt}
\end{tabular}}
\label{table1}
\end{table}

\subsubsection{\textbf{Assessment Score}}
The definition of evaluation items is a central component of an assessment dataset. As noted in prior studies\cite{17,18,19,20,21,22} and in the representative rehabilitation dataset UI-PRMD\cite{32}, many existing approaches treat movements performed by ND as the reference for ideal motion. However, according to Birke et al.\cite{33}, the primary goal of neurological rehabilitation is to improve function and independence. Rehabilitation assessment therefore emphasizes the achievement of functional activities and participation\cite{34}, rather than the mere reduction of impairments. Consequently, even a high-scoring movement performed by a stroke patient should not be directly equated with that of an ND participant, which motivates the need for stroke-specific assessment criteria. Although detailed clinical assessments can be adopted, subjectivity remains a concern. This underscores the necessity of a new set of evaluation items for upper-limb rehabilitation tasks.
The KIMORE\cite{35}\cite{CAPECCI201870} provides an illustrative example of balancing quantitative scoring with clinical interpretation. It employs 10 items rated on a 5-point Likert scale, yielding a maximum score of 50 per trial. Table III summarizes the questionnaire items used in our NRC dataset. We developed evaluation items specifically for the 15 upper-limb movements in our dataset. The items were initially drafted through discussions among three licensed physiotherapists with more than five years of clinical experience and were subsequently validated by two Ph.D researchers in physical therapy. Items 2 and 3 assume that each movement is performed only once, whereas Items 8, 9, and 10 address seated tasks involving upper-limb use. In particular, item 8 evaluates fine motor skills, such as handling small objects, which are essential for ADL.

\subsection{Model Structure}Deb et al.\cite{26} reported that GCN provide an effective framework for assessing rehabilitation exercises. Human body can be formulated as a graph, where skeleton keypoints are treated as nodes and anatomical connections as edges. However, this formulation primarily captures the spatial structure. Because human movements are inherently sequential, the network must also model temporal dynamics. For this reason, ST-GCN have shown strong performance in human action modeling tasks\cite{13}. Rehabilitation movement assessment can be viewed as a subtask of human action analysis and thus benefits from modeling methods, including GCN and ST-GCN. In our approach, a skeleton-based ST-GCN links spatial joint information across time to capture the spatio-temporal characteristics of rehabilitation actions. The overall model architecture is illustrated in Fig. 3. The skeleton representation consists of 25 keypoints, including upper-limb and hand joints. The first ST-GCN block takes as input the skeleton coordinates \(X \in \mathbb{R}^{N \times C \times T \times V \times M}\) and the graph \(G=(V,E)\) with adjacency matrix \(A \in \mathbb{R}^{K \times V \times V}\). Here, \(C\) denotes the channel dimension, \(T\) the sequence length, and \(V\) the number of keypoints. The skeleton graph is constructed for each frame from the joint coordinates, and convolution is performed across both spatial and temporal dimensions. For hop distance \(k\), the adjacency matrix \(A\) is normalized as \(\tilde{A}^{(l)}_{k}=D_{k}^{-1/2}\bigl(A_{k}\odot E_{k}^{(l)}+I\bigr)D_{k}^{-1/2}\). At the \(l\)th layer, each ST\mbox{-}GCN unit is defined as in ~(1) and~(2). Equation~(3) representates an ST-GCN Block.

\begin{equation}
S_l(X) = \sigma\!\left(\sum_{k=1}^{K} \tilde{A}_{k}\, X\, W_{k}^{(l)}\right)
\label{eq1}
\end{equation}

\begin{equation}
\mathit{\Gamma}_{l}(X) = \sum_{c=1}^{C} \sum_{\tau=-\lfloor k_{t}/2\rfloor}^{\lfloor k_{t}/2\rfloor}
U_{C,c,\tau}^{(l)}\, Z_{n,c,\, t + s_{t}\!\cdot\! \tau,\, v} + b_{C}^{(l)}
\label{eq2}
\end{equation}

\begin{equation}
X_l = \sigma\!\big(\mathit{\Gamma}_{l}(S_l(X_{(l-1)})) + r_l(X_{(l-l)})\big)
\label{eq3}
\end{equation}

\begin{figure*}[t]
\centerline{\includegraphics[width=\linewidth]{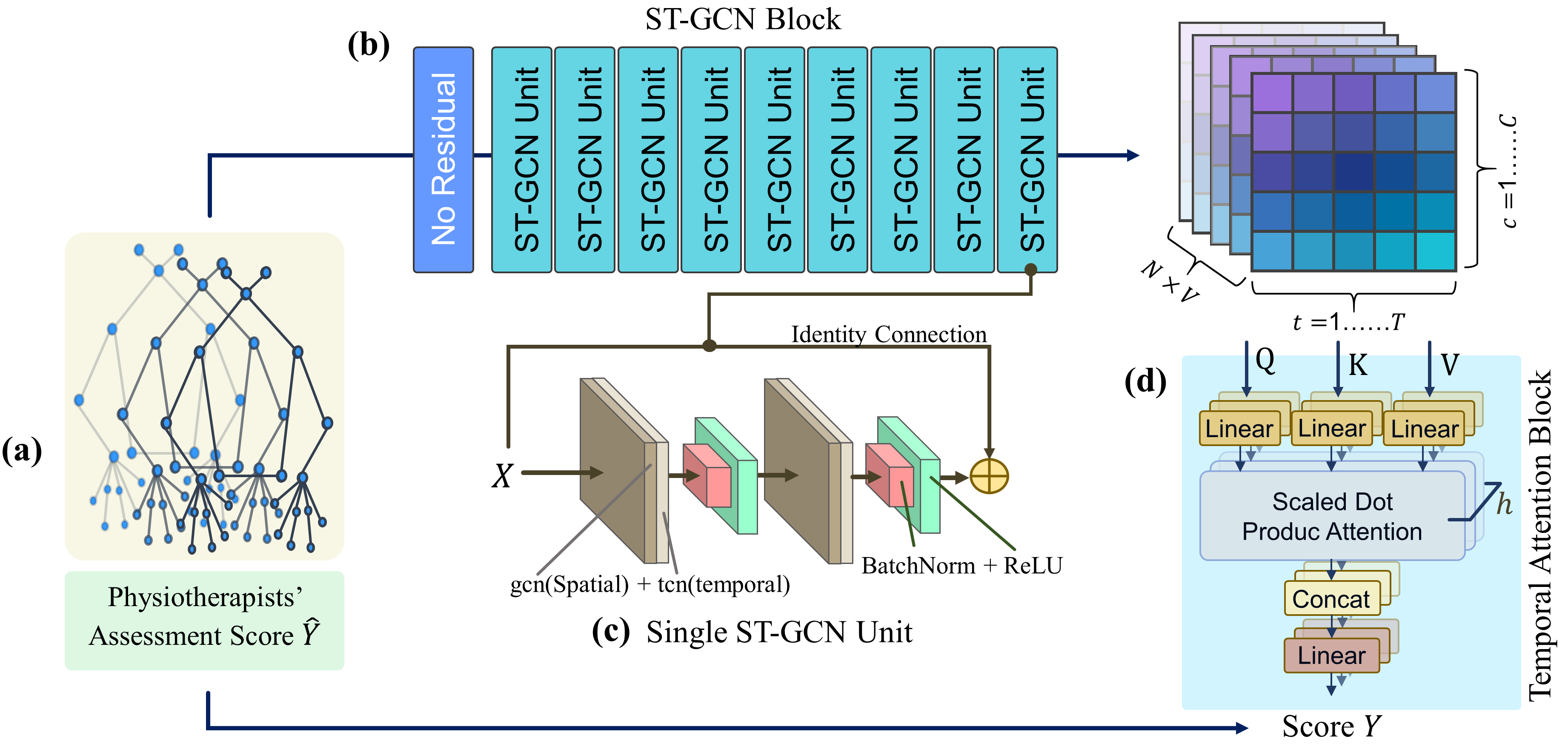}}
\caption{Overview of RAST-G@ model structure.~(a) Skeleton sequence and Score input,~(b) Outputs 3d feature map \(X^{\prime} \in \mathbb{R}^{({N \times V}) \times T \times C}\) via ST-GCN block from skeleton spatio-temporal graph.~(c) Structure of a single ST-GCN unit. After GCN and TCN layers, followed by batch normalization and ReLU; the output is concatenated with the initial input through residual connections.~(d) Temporal Attention(TA) block. From the ST-GCN output, the query (Q), key (K), and value (V) are obtained, and the attention scores are computed as in (4). The prediction \(Y\) is compared with the ground-truth scores, \(\hat{Y}\).}
\label{fig3}
\end{figure*}


Fig.~3(c) illustrates a single ST-GCN unit. All units, except for the first, employ residual connections. After each convolution along the spatial and temporal axes, Batch Normalization and ReLU activation are applied. The output is then combined with the initial input to preserve the original information. While ST-GCN is effective at extracting local spatio-temporal patterns, it treats all joints and temporal segments as equally important. In practical human actions and rehabilitation motions, however, certain joints (e.g., the wrist and shoulder) and specific time points (e.g., the onset or offset of a movement) often contribute more substantially to the overall motion.

Transformers have demonstrated the effectiveness of self-attention mechanisms in deep learning\cite{36}. We incorporate a transformer-based module for temporal feature learning within RAST-G@ to predict assessment scores, as illustrated in Fig.~3(d). Multi-head attention is applied along the temporal axis independently for each joint. This design enables feature learning while performing joint-specific temporal attention, in contrast to LSTM-based models that aggregate temporal information across joints. Because this module does not require additional adjacency-matrix operations, it reduces computational overhead while remaining effective for long-term defendencies. To perform attention along the temporal axis, the Temporal Attention (TA) module reshapes the input \(X\) to \(X^{\prime} \in \mathbb{R}^{({N \times V}) \times T \times C}\) For head \(h\), the query, key, and value matrices are computed as \(Q_h\)=\(X^{\prime}W^Q_h\), \(K_h\)=\(X^{\prime}W^K_h\), and \(V_h\)=\(X^{\prime}W^V_h\). The output is computed as a weighted sum of the values, where the weight assigned to each value is determined by a compatibility function between the query and the corresponding key. The input consists of queries and keys of dimension \(d_k\), and values of dimension \(d_v\). Next, the dot products of the query with all keys are computed, each is scaled by \(\sqrt{d_k}\), and a softmax function is applied to obtain attention weights over the values.\cite{37}


\vspace{-4pt}

\begin{equation}
    Y=W_2 \sigma\!\left(W_1z+b_1\right)+b_2
\label{eq4}
\end{equation}

After learning attention weights over the input sequence, the model outputs a single scalar score representing action quality. A regression head is employed to predict scores across a wide range of conditions. The regression module consists of an average pooling layer followed by a linear layer. The pooling operation aggregates spatio-temporal features into a compact representation \(z^{\prime} \in \mathbb{R}^{N \times C \times 1 \times 2}\), and the linear layer maps this representation to a scalar score \(Y\), as in~(4). The final model output \(Y\) is compared with the ground-truth scores assigned by expert physiotherapists. For model optimization, the \textit{Huber Loss},~(5) is used with delta($\delta$)=0.1, which combines the properties of MSE and MAE. When the prediction error is smaller than $\delta$, the loss behaves like MSE and is sensitive to small deviations. When the error exceeds $\delta$, it behaves like MAE, thereby reducing the influence of outliers.

\vspace{-2pt}
\begin{equation}
 L_{\delta}(y, \hat{y}) =
    \begin{cases}
    \mathrm{MSE}(y-\hat{y}), & \lvert y-\hat{y}\rvert \le \delta \\[2pt]
    \delta\,\mathrm{MAE}(y-\hat{y}) - \dfrac{\delta^{2}}{2}\, & \lvert y-\hat{y}\rvert > \delta
    \end{cases}
\label{eq5}
\end{equation}

\section{Experiment}
\subsection{Setup}
\subsubsection{\textbf{Dataset}}
We conduct experiments on datasets scored on a 50-point Likert-based scale. The evaluated datasets include both whole-body and upper-limb rehabilitation exercises. Whole-body tasks reflect global motor coordination, whereas upper-limb tasks emphasize fine motor control of the arms and hands. By conducting experiments on these two datasets, we assess the model performance across different types of rehabilitation actions. \textbf{a) KIMORE} Kinematic Assessment of Movement for Remote Monitoring of Physical Rehabilitation (KIMORE)\cite{35} is a dataset comprising RGB-D videos and score annotations for five whole-body exercises. Participants are grouped according to the presence or absence of pain or disability. The control group (CG) includes movements performed by 12 professionals, including physiotherapists, and 32 non-professionals. The group with pain and postural disorders(GPP) consists of 34 participants with conditions such as Parkinson’s disease, back pain, stroke. For each exercise, scores were provided by a professional rehabilitation therapist.

\textbf{b) NRC(ours)} This dataset consists of upper-limb rehabilitation movement data collected using RGB-D cameras, IMU sensors, and dedicated software.\cite{ye-eun} It includes 15 exercises comprising 10 ADL and 5 ROM tasks, as listed in Table II. The ADL set emphasizes upper-limb and finger movements relevant to daily life. We further categorized the exercises into three groups: UNI (unimanual hand use), BIA (bimanual asymmetric hand use), and BIS (bimanual symmetric hand use). These categories enable comparative analysis of model performance in mapping diverse action types to assessment scores. 
All action samples in both datasets are annotated with scores provided by professional rehabilitation therapists. For NRC, we developed a new set of 10 dedicated assessment items. The assessment score is derived from these items.

We designed an end-to-end deep learning system for rehabilitation action recognition and assessment, where the proposed model serves as the assessment module. For this task, the model was trained on data collected from 6 ND participants and 11 stroke patients. The effectiveness of rehabilitation depends on the time since stroke onset\cite{51} and repetitive practice. This consideration motivates the primary goal of our system: enabling patients to perform rehabilitation exercises independently at home. In particular, participant recruitment focused on subacute stroke with hemiplegia within six months of stroke onset. The detailed inclusion criteria for the recruited stroke participants are as follows:

\vspace{0.5\baselineskip}
\begin{itemize}
	\item Age 19 years or older
	\item Subacute or chronic stage
	\item Upper-Limb hemiplegia
	\begin{itemize}
		\item FMA $\geq$ 30
        \item Brunnstrom's stage hand recovery 4-4.5 stage
    \end{itemize}
	\item Understand and communicate about the experiment
	\begin{itemize}
		\item MOCA $\geq$ 22 
    \end{itemize}
\end{itemize}
\vspace{0.5\baselineskip}

In total, 1,142 movement–score pairs were used, segmented per exercise. The dataset was split into train, validation, and test sets at approximately 8:1:1 ratio. A summary of the datasets is provided in Table I.

\subsubsection{\textbf{Preprocessing}}
\textbf{a) Pose estimation} The raw data consist of RGB-D video sequences, which require an additional skeleton keypoint estimation step. For this purpose, we employed MMPose, an open-source human pose estimation library developed by the OpenMMLab team.\cite{mmpose} This framework provides a unified interface for training, inference, and evaluation across a wide range of pose estimation models. In our work, we ran inference using the HRNet-DarkPose model\cite{39,40}, trained on the COCO-WholeBody dataset\cite{31}, to estimate human poses. The extracted skeleton representation is presented in Fig. 1.
\textbf{\text{b)} Uniform Frame Sequences}
Standardizing frame sequences into a uniform format is a critical preprocessing step for deep learning models. Although individual actions may differ in temporal characteristics such as duration and speed, consistency across the dataset is necessary. Inevitably, this process involves partial information loss. To mitigate this, we applied a frame-dropping preprocessing strategy. As illustrated in Fig. 4, the original sequence was divided into non-overlapping groups of $N$ frames. One frame was then sampled from each group and concatenated in temporal order. This approach preserves the overall temporal trend of the action.
It also enables random frame selection within each group, which increases data diversity and facilitates augmentation. Using this strategy, all sequences were standardized to a length of 288 frames. The raw data distribution is reported in another study of ours.\cite{ye-eun}

\begin{figure}[t]
\centerline{\includegraphics[width=\linewidth]{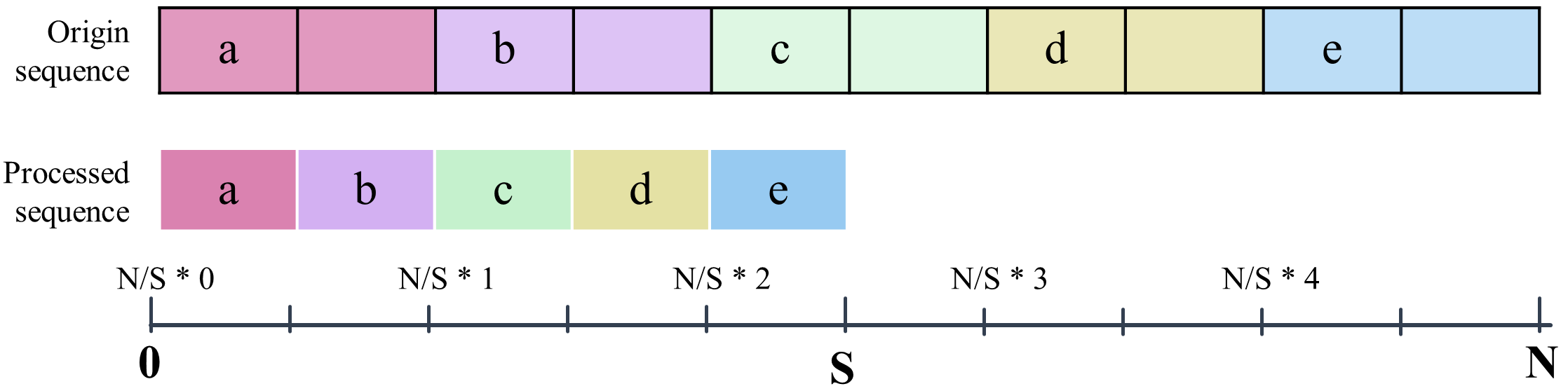}}
\caption{Uniform frame sequence method by frame dropping.}
\label{fig4}
\end{figure}

\subsection{Training}
The implementation and training were conducted on an Intel i9-10920X CPU and an NVIDIA GeForce RTX 4090 GPU. All models were implemented in PyTorch and trained on Ubuntu 22.04.5 LTS. The hyperparameters were set as follows: 200 epochs, a batch size of 64, and the AdamW optimizer with a learning rate of 0.003. For performance evaluation, Mean Absolute Deviation (MAD), Root Mean Square Error (RMSE), and Mean Absolute Percentage Error (MAPE) were employed. The equations are as follows:

\begin{equation}
\text{MAD} = \frac{1}{n} \sum_{i=1}^{n} \left| y_i - \hat{y}_i \right|
\label{eq6}
\end{equation}

\begin{equation}
\text{RMSE} = \sqrt{ \frac{1}{n} \sum_{i=1}^{n} \left( y_i - \hat{y}_i \right)^2 }
\label{eq7}
\end{equation}

\begin{equation}
\text{MAPE} = \frac{1}{n} \sum_{i=1}^{n} \left| \frac{y_i - \hat{y}_i}{y_i} \right| \times 100
\label{eq8}
\end{equation}

Each metric is computed over the entire test set and reported as an average error between ground-truth scores and model predictions.

\section{Result}

\begin{table*}[t]
  \centering
  \caption{COMPARISON OF ASSESSMENT DEEP LEARNING MODEL RESULTS ON KIMORE USING RMSE, MAPE AND MAD METRICS (LOWER VALUES INDICATE BETTER PERFORMNACE)}
  \label{tab:kimore}

  \footnotesize
  \setlength{\fboxsep}{15pt}
  \setlength{\arrayrulewidth}{0.5pt}
  \setlength{\tabcolsep}{6pt}   
  \renewcommand{\arraystretch}{1.15}
    \begin{tabular}{c c *{4}{c} *{5}{c}}
    \noalign{\hrule height 0.8pt}
    \hline
     & & \multicolumn{4}{c}{\raisebox{-\height}[0pt][0pt]{Assessment}} & \multicolumn{5}{c}{\raisebox{-\height}[0pt][0pt]{Recognition + FC Layer}}\\

    Metric & Ex &
    \multicolumn{4}{c}{{\rule{6.25cm}{0.5pt}}} &
    \multicolumn{5}{c}{{\rule{7.5cm}{0.5pt}}} \\
     &  & \shortstack{\textbf{RAST-G@}\\(ours)} & \shortstack{Kuang \textit{et al}. \\ \cite{41}} & \shortstack{Deb \textit{et al}. \\ \cite{26}} & \shortstack{Liao \textit{et al}. \\ \cite{23}} &
    \shortstack{Zhang \textit{et al}. \\ \cite{42}} & \shortstack{Song \textit{et al}. \\ \cite{43}} & \shortstack{Yan \textit{et al}. \\ \cite{13}} & \shortstack{Li \textit{et al}. \\ \cite{44}} & \shortstack{Du \textit{et al}. \\ \cite{45}} \\
    \hline
        & Ex1 & \textbf{0.267} & 0.399 & 2.024 & 2.534 & 2.916 & 2.165 & 2.017 & 2.344 & 2.440 \\
        & Ex2 & \textbf{0.268} & 0.354 & 2.120 & 3.738 & 4.140 & 3.345 & 3.262 & 2.823 & 4.297 \\
    \raisebox{-\height}[0pt][0pt]{RMSE $\downarrow$} & Ex3 & \textbf{0.274} & 0.289 & 0.556 & 1.561 & 2.615 & 1.929 & 0.799 & 2.004 & 1.925 \\
        & Ex4 & 0.264 & \textbf{0.101} & 0.644 & 0.792 & 1.836 & 2.018 & 1.331 & 1.078 & 1.676 \\
        & Ex5 & \textbf{0.264} & 0.386 & 1.181 & 1.914 & 2.916 & 3.198 & 1.951 & 2.575 & 3.158 \\
    \cline{2-11}
        & AVG & \textbf{0.267} & 0.306 & 1.305 & 2.108 & 2.8846 & 2.531 & 1.872 & 2.165 & 2.699 \\
    \hline
        & Ex1 & \textbf{0.364} & 0.431 & 1.926 & 2.589 & 5.054 & 2.605 & 2.339 & 3.491 & 3.228 \\
        & Ex2 & \textbf{0.370} & 0.749 & 1.272 & 3.976 & 10.436 & 3.296 & 6.136 & 5.298 & 6.001 \\
    \raisebox{-\height}[0pt][0pt]{MAPE $\downarrow$} & Ex3 & 0.385 & \textbf{0.271} & 0.728 & 2.023 & 5.774 & 2.968 & 1.727 & 4.188 & 3.421 \\
        & Ex4 & 0.346 & \textbf{0.173} & 0.824 & 2.333 & 3.901 & 2.152 & 2.325 & 1.976 & 2.584 \\
        & Ex5 & \textbf{0.349} & 0.692 & 1.591 & 2.312 & 6.531 & 4.959 & 3.802 & 5.752 & 5.620 \\
    \cline{2-11}
        & AVG & \textbf{0.363} & 0.463 & 1.268 & 2.647 & 6.3392 & 3.196 & 3.266 & 4.141 & 4.171 \\
    \hline
        & Ex1 & 0.225 & \textbf{0.186} & 0.799 & 1.141 & 1.757 & 0.977 & 0.889 & 1.378 & 1.271 \\
        & Ex2 & \textbf{0.227} & 0.235 & 0.774 & 1.528 & 3.139 & 1.282 & 2.096 & 1.877 & 2.199 \\
    \raisebox{-\height}[0pt][0pt]{MAD $\downarrow$} & Ex3 & 0.231 & \textbf{0.111} & 0.369 & 0.845 & 1.737 & 1.105 & 0.604 & 1.452 & 1.123 \\
        & Ex4 & 0.221 & \textbf{0.053} & 0.347 & 0.468 & 1.202 & 0.715 & 0.842 & 0.675 & 0.880 \\
        & Ex5 & \textbf{0.220} & 0.223 & 0.621 & 0.947 & 1.853 & 1.536 & 1.218 & 1.662 & 1.864 \\
    \cline{2-11}
        & AVG & 0.225 & \textbf{0.162} & 0.582 & 0.986 & 1.938 & 1.123 & 1.130 & 1.409 & 1.467 \\
    \noalign{\hrule height 0.75pt}
  \end{tabular}
\end{table*}

\subsection{Overview}
We carefully selected and categorized baseline models for comparison. Although deep learning–based rehabilitation assessment remains a promising research area, the number of well-validated prior studies is still limited. Accordingly, we adopted established human action recognition models as baselines. The models listed in Table IV under the Recognition Model–FC layer category correspond to this setting. The methods\cite{13,42,43,44,45} were originally designed for recognition tasks; for assessment, we replaced their softmax based classification heads with fully connected layers to perform score regression. This adaptation is feasible because both recognition and assessment rely on extracting discriminative features from human action data.
Because human behavior is a high-level data that can be captured through multiple modalities, several studies have explored multimodal deep learning rather than relying on a single modality. For example, Kuang et al.\cite{41} proposed a hierarchical approach that separately learns position and orientation and then integrates them to form a joint representation.

\subsubsection{KIMORE Dataset Result} We compared our model, RAST-G@, with baseline models using the three evaluation metrics defined in (6), (7), (8). As reported in Table IV, the proposed model achieved lower errors and maintained consistent performance across all five exercises. These results suggest that the model is applicable not only to whole-body tasks but also to a broader range of rehabilitation movements and assessment settings. Among the three metrics, MAD showed only modest improvement over prior models, whereas RMSE and MAPE exhibited substantially larger gains. This difference can be attributed to how the metrics summarize prediction errors. MAD averages absolute deviations and is less sensitive to large errors, while RMSE and MAPE more strongly reflect the impact of large deviations and their relative magnitude. 
Regarding Ex4, the results can be explained by its motion characteristics. Prior work has reported particularly low errors for this exercise, which is consistent with its motion characteristics. Ex4 mainly involves pelvic rotation in the transverse plane while other body parts remain relatively fixed. In this setting, rotational variation may be more informative than translational displacement. Under a Cartesian representation, z-axis can dominate the motion signal, whereas quaternion-based representations can capture rotational components more directly. Consequently, the method of Kuang et al. is well suited for Ex4 and yields very small errors.

Overall, these results indicate that our model is not specialized for a single exercise but instead maintains robust performance across a diverse set of actions. This robustness becomes more evident as the number of evaluated exercises increases and inter-exercise variability grows. Accordingly, RAST-G@ is advantageous for stroke rehabilitation assessment, where patient-specific movement patterns vary substantially and long-term monitoring is required.



\begin{table}[ht]
\caption{COMPARISON OF ASSESSMENT DEEP LEARNING MODEL RESULTS ON NRC DATASET(OURS)}
\label{table4}
\centering
\setlength{\tabcolsep}{10pt}
{\renewcommand{\arraystretch}{1.25}
    \begin{tabular}{c c c c}
        \noalign{\hrule height 0.75pt}
        \hline
        {\centering Metric\par} & {\centering \textbf{RAST-G@}\par} & {\centering Kuang \textit{et al}.\cite{41}\par} & {\centering Deb \textit{et al}.\cite{26}\par} \\
        \hline
        RMSE $\downarrow$ & \textbf{0.291} & 1.961 & 1.030\\
        MAPE $\downarrow$ & \textbf{0.259} & 2.836 & 3.805\\ 
        MAD $\downarrow$ & \textbf{0.321} & 0.739 & 0.892\\
        \hline
        \noalign{\hrule height 0.75pt}
    \end{tabular}}
\end{table}

\subsubsection{NRC(Ours) Dataset Result}
The NRC dataset collected in this study includes both ADL and ROM exercises. ADL tasks involve relatively complex and fine-grained movements and may be more challenging than other actions, depending on stroke severity. Moreover, they are limited to the upper extremities, such that motion information are concentrated in the arms and fingers.
For baseline comparisons, we evaluated our method against graph-based deep learning approaches proposed by Kuang et al.\cite{41} and Deb et al.\cite{26}. Kuang et al. employed a hierarchical strategy that learns position and orientation separately and then integrates them into a joint representation. Orientation was represented using four-dimensional quaternions \((x, y, z, w)\), which encode the directional state of each joint.\cite{46}
Accordingly, we converted the Cartesian skeleton coordinates into a quaternion-based representation prior to experimentation. Given a skeleton sequence  \(seq\) and the skeletal connectivity defined by parent–child joint relations, we first computed a reference REST pose to serve as the baseline for rotation estimation. For each frame, joint orientations were estimated relative to this baseline by comparing each parent–child bone direction with the corresponding direction in the REST pose. To obtain a stable and consistent rotation parameterization, a reference vector \(up_0\) was introduced to determine the roll component around the \(y\)-axis and to enforce orthogonality among the \(x, y\), and \(z\)-rotation axes. Repeating this process over all frames produces a sequence of joint-wise quaternions. The resulting orientation sequence is represented as \(Q \in \mathbb{R}^{T \times J \times 4}\), where each element corresponds to a quaternion describing the orientation of joint \(j\) at time \(t\).
When trained using both position and orientation features, RAST-G@ showed substantial improvements over the baseline models on the NRC dataset. NRC contains target actions that require fine finger control and exhibit temporal structure, which allows the proposed temporal modeling strategy to be particularly effective. As summarized in Table V, while other baselines exhibited up to a fivefold increase in error when transferred to NRC, RAST-G@ maintained stable performance, demonstrating robustness under dataset shift. Moreover, RAST-G@ consistently outperformed the baselines on error measures that emphasize large deviations and relative error magnitude. Notably, MAPE was lower than the baselines, indicating robustness across subjects and task types.
Fig. 5 presents the ground-truth scores provided by therapists and the corresponding predictions of RAST-G@ on the NRC dataset. A longitudinal study was conducted monthly over four months, during which movement data and therapist-assigned scores were collected for assessment. The comparison indicates that the model captures the overall trend in scores and predicts accurate score.

\begin{figure}[t]
\centerline{\includegraphics[width=\linewidth]{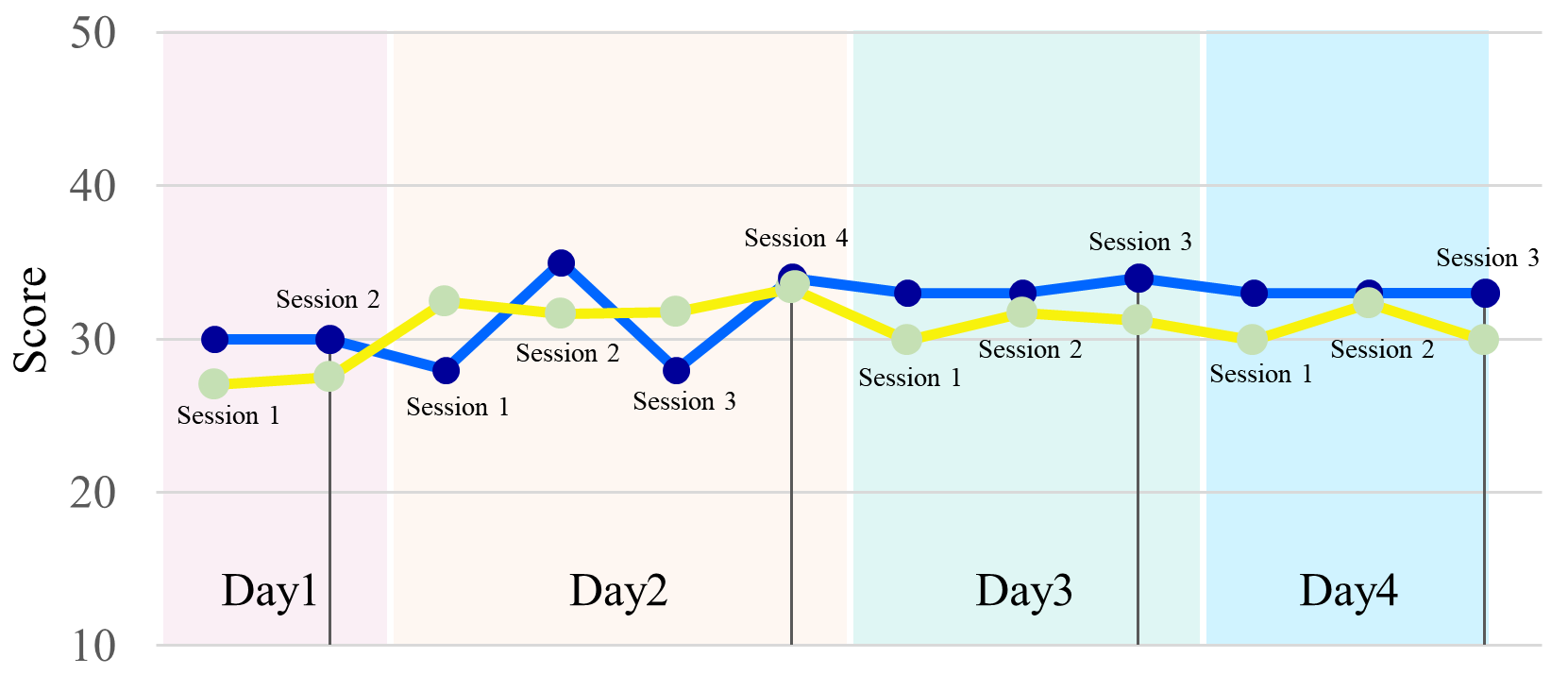}}
\caption{Visualization of NRC works(yellow) compare to theraphists' score(blue) during 4 month long-term dataset. Label02: brushing hair, Stroke 14.}
\label{fig5}
\end{figure}

\subsection{Ablation Study}
The ablation study in Table VI provides a comparison of the effect of each method. The frame standardization strategy used to unify input can affect the model’s predictions. Frame dropping reduces sequence length while preserving the overall temporal progression. Without frame dropping, MAPE exhibited a increase, whereas MAD and RMSE remained unchanged. This indicates inadequate temporal preservation during standardization distorts action features and increases prediction error.

The TA block was introduced to capture temporal dependencies in rehabilitation movements. Without the TA block caused the largest performance degradation, highlighting the importance of temporal information when mapping motions to qualitative assessment scores. Overall, this indicates that temporal modeling is essential for understanding rehabilitation tasks and more broadly, human actions.

\begin{table}[t]
\caption{Method Ablation study result}
\label{table5}
\centering
\setlength{\tabcolsep}{10pt}
{\renewcommand{\arraystretch}{1.25}
    \begin{tabular}{c c c c}
        \noalign{\hrule height 0.75pt}
        \hline 
        {\centering Method\par} & {\centering MAD $\downarrow$\par} & {\centering RMSE $\downarrow$\par} & {\centering MAPE $\downarrow$\par} \\
        \hline
        w/o Frame Drop & 0.260 & 0.326 & 0.379\\
        w/o Temporal Attention(TA) & 0.277 & 0.340 & 0.480\\ 
        Keypoints: 27 & 0.261 & 0.318 & 0.427\\
        Stroke Only & 0.250 & 0.303 & 0.335\\ 
        \hline
        \textbf{Proposal} & 0.259 & 0.321 & 0.291\\
        \hline
        \noalign{\hrule height 0.75pt}
    \end{tabular}}
\end{table}

\begin{figure}[b]
\centerline{\includegraphics[width=\linewidth]{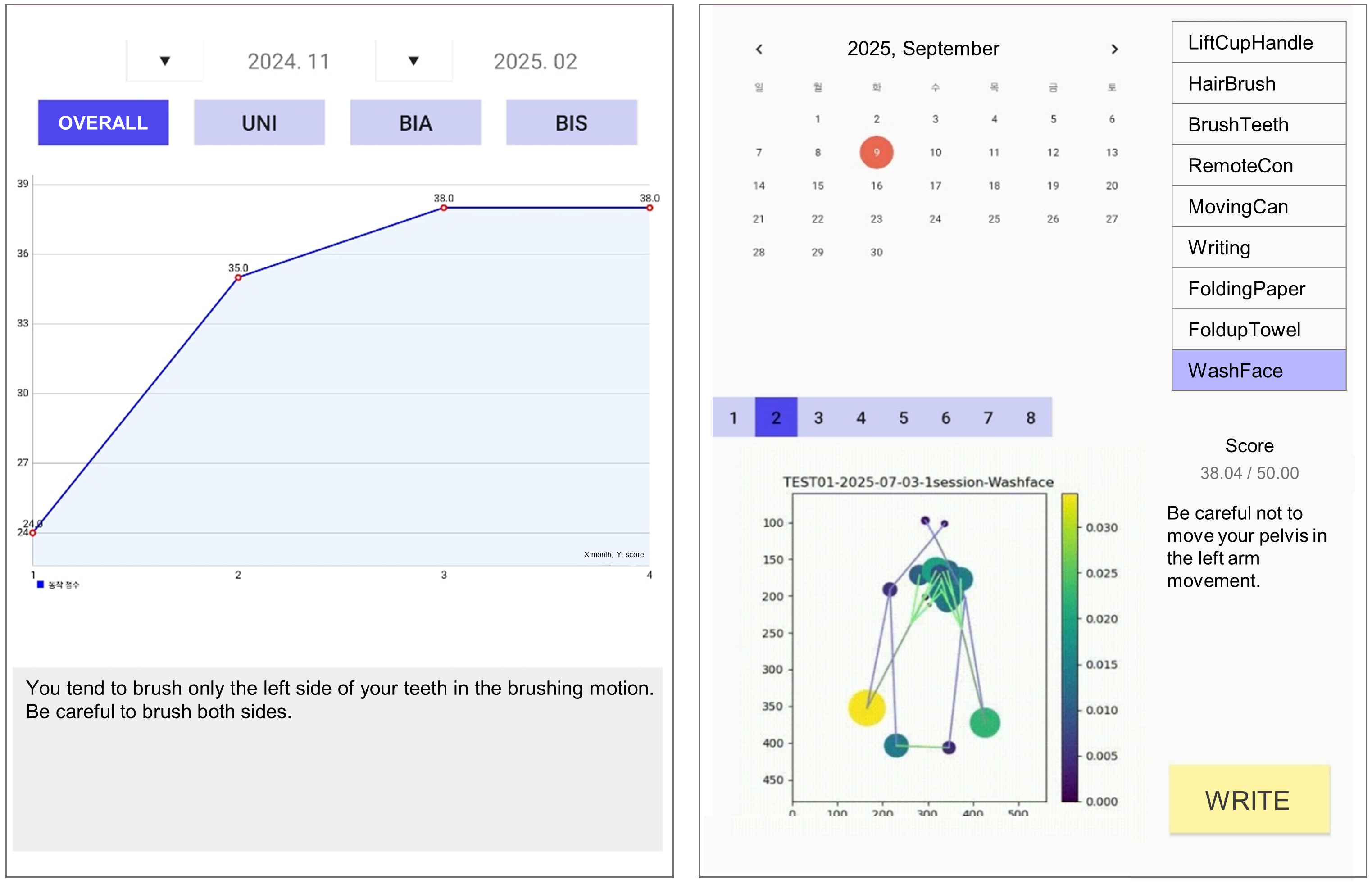}}
\caption{Feedback from our application, Period feedback offers graph during exercise periods(left), Discrete feedback shows skeleton graph heatmap, score and comment each per exercise.(right)}
\label{fig6}
\end{figure}

The Keypoint 27 setting was introduced as a comparison factor for optimizing data representation, rather than for validating the proposed method itself. As shown in Fig. 1, we derived skeleton from the COCO WholeBody configuration. However, our target movements differ from those in the benchmark datasets used for pose estimation. We therefore evaluated whether the skeleton is appropriate for our data.
Prior work has emphasized the value of selecting informative landmarks. \cite{47, 48} introduced attention mechanisms to focus on important keypoints in full-body motion, and related efforts in sign-language modeling address dense hand regions. \cite{49} used a skeleton graph with a hand-focused structure  reporting a drop in accuracy to 63.69 when apply graph node reduction. Since our dataset includes ADL tasks involving fine-grained hand motions, we evaluated a comparable configuration and compared the resulting performance.
However, MAPE decreased, indicating improved relative accuracy. This suggests that, for human action tasks requiring holistic motion understanding, the choice of skeleton representation should prioritize consistency over increased dimensionality. For sequential action data with substantial temporal and spatial complexity, reducing redundant information while preserving overall structure can be more effective.

Unlike reference-based approaches\cite{16,17,18,19,20,22} that assign scores by comparing an input sequence to an ideal movement template, our method learns a direct mapping from action patterns to assessment scores using supervised training data. As shown in the stroke-only ablation results, training on stroke data alone achieved performance comparable to the proposed setting. Because both training and testing were performed on stroke sequences, absolute-error metrics such as MAD remained similar and in some cases slightly improved. In contrast, RMSE and MAPE were higher than in the ND–stroke setting, suggesting that including ND sequences helps the model capture smoother movement flow, whereas stroke sequences often contain more abrupt temporal variations. Overall, RAST-G@ with optimized frame construction, the TA module, and a 25-keypoint skeleton maintained low MAD and RMSE while reducing MAPE. The model not only limits absolute error but also improves accuracy with respect to motion magnitude, which is important for rehabilitation assessment and broader human action understanding.

\subsection{Feedback for User}
In designing a domiciliary rehabilitation system, we also considered the need for user-friendly feedback. Model outputs should be interpretable not only to clinicians but also to patients and older adults who may have limited familiarity with artificial intelligence, computer systems, or medical terminology. Based on this requirement, we developed a mobile application that provides structured feedback to users as shown in Fig. 6. The feedback is organized into two components: period feedback and discrete feedback.
\subsubsection{Period feedback} Period feedback summarizes performance trends over time. For each month, a representative score computed as an average is reported to visualize changes in rehabilitation progress. The time window can be adjusted according to user preference, and performance graphs are available for both overall scores and the categories OVERALL, UNI, BIA, and BIS. Because task difficulty differs across categories, this view enables more fine-grained monitoring. Such longitudinal feedback can also support therapists in designing and adjusting individualized treatment plans.
\subsubsection{Discrete feedback}
Discrete feedback records the outcome of each exercise by providing skeleton heatmaps, task-specific scores, and, when available, therapist comments. This feedback offers patients an intuitive summary of performance and enables clinicians to document qualitative observations over time. Fig. 7 presents the vizualization feedback interface of the proposed system.
The visual feedback is generated from the assessment module, including RAST-G@, and highlights the most influential joints and edges in the motion sequence \(m\) at frame \(t_1\). These regions are displayed as heatmaps overlaid on the skeleton sequence, where color intensity and node size indicate their contribution to the assessment.
The input tensor \(X \in \mathbb{R}^{N \times C \times T \times V \times 1}\) is processed by ST-GCN blocks to capture skeletal structure and temporal context while preserving the skeleton configuration. The resulting feature representation \(X^L \in \mathbb{R}^{{(N \times M)} \times 256 \times {(T/4)} \times V}\) is then used to generate skeleton heatmaps for visual feedback. For clarity, the score is rescaled to a 0–100 range.

In beta testing with physiotherapists, the interface was well received, particularly for visualizing postural alignment and joint utilization. Beyond improving usability, this feedback enables analysis of how the model attends to different body regions during inference, supporting interpretability and informing future model analysis and refinement.

\begin{figure}[t]
\centerline{\includegraphics[width=0.8\linewidth]{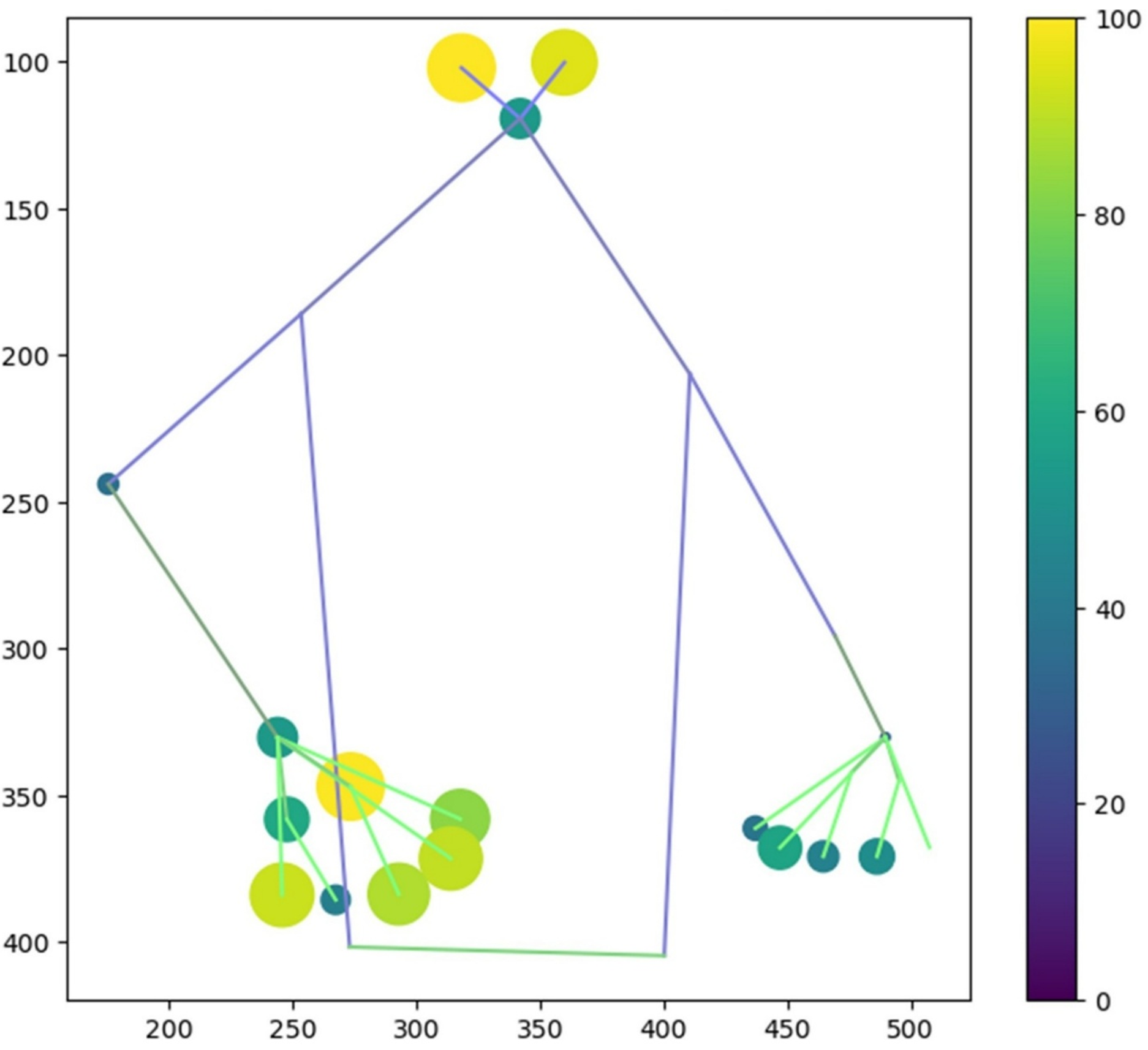}}
\caption{Feedback with skeleton graph heatmap. Subject Stroke13, Label07: folding paper.}
\label{fig5}
\end{figure}

\section{Conclusion}
Our approach provides assessment scores without requiring a non-disabled control group or reference motions. Collecting rehabilitation movement data is costly for both impaired and non-impaired participants. Moreover, rehabilitation assessment focuses on functional activity and participation. Accordingly, our method supports patient-centered evaluation by analyzing each patient’s own movements and performance without relying on non-patient reference data.

Nevertheless, the proposed model (RAST-G@) is limited to assessment and does not yet address treatment planning within a comprehensive home-based rehabilitation system. Future work will focus on an integrated higher-level framework that can generate personalized recovery plans based on more detailed analyses of individual exercises.


\section*{Acknowledgment}
This study was supported by the Translational Research Center for Rehabilitation Robots(\#NRCTR-EX23008), National Rehabilitation Center, Ministry of Health and Welfare, Korea.
and the Gachon University research fund of 2023(GCU-202300710001).

\bibliographystyle{IEEEtran}
\bibliography{ref}

\end{document}